\documentclass{llncs}

\usepackage{graphicx}
\usepackage{url}
\urldef{\mailsa}\path|{hubg@nlpr.ia.ac.cn}|
\usepackage{hyperref}
\usepackage{CJKutf8}
\usepackage{indentfirst}
\begin{document}

\title{Information Theory and its Relation to \\ Machine Learning}

\author{Bao-Gang Hu}

\institute{National Laboratory of Pattern Recognition, Institute of Automation, \\
Chinese Academy of Sciences, Beijing 100190, China \\
\mailsa\\
}
\maketitle

\begin{abstract}

In this position paper, I first describe a new perspective on machine learning (\textbf{ML}) by 
four basic problems (or levels), namely, {\it ``What to learn?''}, {\it ``How to learn?''}, {\it ``What to evaluate?''}, 
and {\it ``What to adjust?''}. The paper stresses more on the first level of {\it ``What to learn?''}, 
or {\it ``Learning Target Selection''}. Towards this primary problem within the four levels, 
I briefly review the existing studies about the connection between information theoretical learning 
(\textbf{ITL} \cite{Principe}) and machine learning. 
A theorem is given on the relation between the empirically-defined similarity measure and 
information measures. Finally, a conjecture
is proposed for pursuing a unified mathematical interpretation to learning target selection. 

\end{abstract}

\begin{quotation}
\noindent {\it ``From the Tao comes one, from one comes two, from two comes three, and from three comes all things.''} \cite{Lao}
\begin{flushright} - by Lao Tzu (ca. 600-500 BCE) \end{flushright}
\indent {\it ``Nature is the realization of the simplest conceivable mathematical
ideas.''} \cite{Einstein} 
\begin{flushright} - by Albert Einstein (1879-1955) \end{flushright}
\end{quotation}

\section{Introduction}\label{sec:Introduction}

\noindent Machine learning is the study and construction of systems that can learn from data.
The systems are called {\it learning machines}. 
When Big Data emerges increasingly, more learning machines are
developed and applied in different domains.
However, the ultimate goal of machine learning study
is {\it insight}, not machine itself. By the term insight I mean {\it learning mechanisms} in descriptions of mathematical principles.
In a loose sense, learning mechanisms can be regarded as the natural entity.
As the {\it ``Tao}\begin{CJK}{UTF8}{bsmi}
(道)\end{CJK}{''} reflects the most fundamental of the universe by Lao Tzu \begin{CJK}{UTF8}{bsmi}(老子)\end{CJK}, 
Einstein suggests that we should pursue the simplest mathematical 
interpretations to the nature. 
Although learning mechanisms are related to the subjects of psychology, cognitive and brain science,
this paper stresses on the exploration of mathematical principles for interpretation of learning mechanisms.
Up to now, we human beings are still far away from deep understanding ourself on
learning mechanisms in terms of mathematical principles. 
It is the author's belief that  {\it ``mathematical-principle-based machine''}  
might be more important and critical than  {\it ``brain-inspired machine''}  
in the study of machine learning. 

The purpose of this position paper is to put forward a new perspective and
a novel conjecture within the study of machine learning. 
In what follows I will present four basic problems (or levels) in machine learning.
The study on information theoretical learning is briefly reviewed. 
A theorem between the empirically-defined similarity measures 
and information measures are given. Based on the existing investigations,
a conjecture is proposed in this paper. 

\section{Four basic problems (or levels) in machine learning}\label{sec:Second}

For information processing by a machine, in the 1980's, Marr \cite{Marr} proposed 
a novel methodology by three distinct yet complementary levels, namely, {\it ``Computational theory''}, 
{\it ``Representation and algorithm''}, and {\it ``Hardware implementation''}, respectively. 
Although the three levels are {\it ``coupled''} loosely, the distinction is of great necessity to 
isolate and solve problems properly and efficiently. In 2007, Poggio \cite{Poggio} described 
another set of three levels on learning, namely, {\it ``Learning theory and algorithms''}, 
{\it ``Engineering applications''}, and {\it ``Neuroscience: models and experiments''},  respectively. 
In apart from showing a new perspective, one of important contributions of this methodology 
is on adding a {\it closed loop} between the levels. These studies are enlightening because 
they show that complex objects or systems should be addressed by decompositions with different, 
yet basic, problems. The methodology is considered to be {\it reductionism} philosophically. 

In this paper, I propose a novel perspective on machine learning by four levels shown in Fig. 1. 
The levels correspond to four basic problems. The definition of each level is given below.

\begin{figure}[!htb]
    \center{
    \hspace*{2cm} \includegraphics[width=55mm]{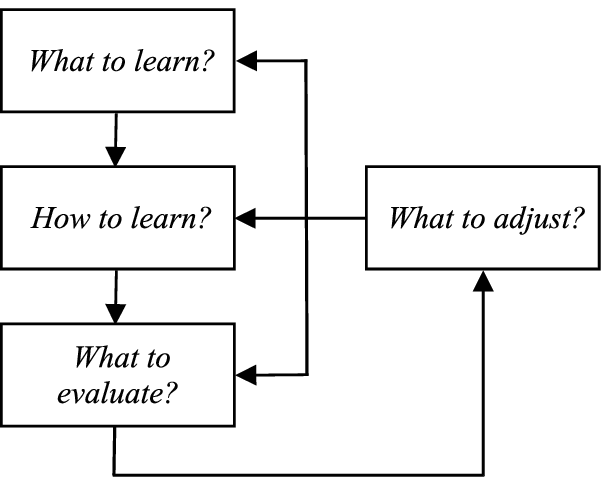}  
    \newline
    \small Fig. 1. Four basic problems (or levels) in machine learning.
}
\end{figure}

\textbf{Definition 1:} 
{\it ``What to learn''} is a study on identifying learning target(s) to the given
problem(s), which will generally involve
two distinct sets of representations (Fig. 2) defined below. 

\textbf{Definition 1a:} 
{\it ``Linguistic representation''} reflects a {\it high-level} description in a {\it natural language} about the expected learning information. This study is more related to linguistics, psychology, and cognitive science. 

\textbf{Definition 1b:} 
{\it ``Computational representation''} 
is to define the expected learning information based on {\it mathematical notations}. It is relatively a {\it low-level} representation which generally includes objective functions, constraints, and optimization formations.

\textbf{Definition 2:} 
{\it ``How to learn?''}
is a study on learning process design and implementations. Probability, statistics, utility, optimization, 
and computational theories will be the central subjects. 
The main concerns are generalization performance, robustness, model complexity, 
computational complexity/cost, etc. 
The study may include physically realized system(s).

\textbf{Definition 3:} 
{\it ``What to evaluate?''}
is a study on {\it ``evaluation measure selection''} where evaluation measure
is a mathematical function. 
This function can be the same or different with the objective function defined in the first level. 

\textbf{Definition 4:} 
{\it ``What to adjust?''}
is a study on dynamic behaviors of a machine from adjusting its component(s). 
This level will enable a machine with a functionality of {\it ``evolution of intelligence''}.

\begin{figure}[htb]
    \centering{
    \hspace*{1.9cm}  \includegraphics[width=50mm]{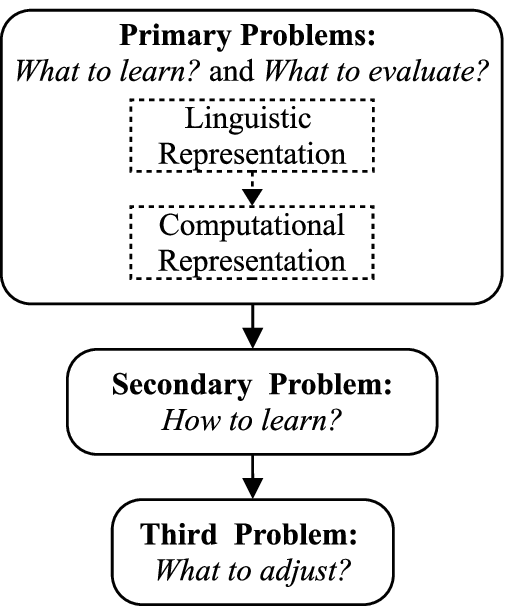}  \newline
    \small Fig. 2. Design flow according to the basic problems in machine learning.
}
\end{figure}

The first level is also called {\it ``learning target selection''}.
The four levels above are neither mutually exclusive, nor collectively exhaustive to every problems in machine learning. 
We call them {\it basic} so that the extra problems can be merged within one of levels.
Figs. 1 and 2 illustrate the relations between each level in different contexts, respectively. 
The problems within four levels are all inter-related, particularly for {\it ``What to learn?''} 
and {\it ``What to evaluate?''} (Fig. 2). 
{\it ``How to learn?''} may influence to {\it ``What to learn?''}, such as convexity of the objective function
or scalability to learning algorithms \cite{Bengio2007}
from a computational cost consideration. 
Structurally, {\it ``What to adjust?''} level is applied to provide the multiple closed 
loops for describing the interrelations (Fig. 1). 
Artificial intelligence will play a critical role via this level.
In the {\it ``knowledge driven and data driven”} model \cite{Hu2009}, 
the benefits of utilizing this level are shown from 
the given examples by {\it removable singularity hypothesis} to {\it ``Sinc''} function and {\it prior updating} 
to Mackey-Glass dataset, respectively.
Philosophically, {\it ``What to adjust?''} level remedies the intrinsic problems in the methodology of {\it reductionism} and 
offers the functionality power for being {\it holism}. However, 
this level receives even less attention 
while learning process holds a {\it self organization} property. 

I expect that the four levels show a novel perspective about 
the basic problems in machine learning. Take an example shown in Fig. 3 
[after Duda, et al, \cite{Duda}, Fig. 5-17] . Even for 
the linearly separable dataset, the learning function using least mean square (\textbf{LMS}) does 
not guarantee a  {\it ``minimum-error''} classification. This example demonstrates two points. 
First, the computational representation of LMS is not compatible with the linguistic representation 
of {\it ``minimum-error''} classification. 
Second, whenever a learning target is wrong in the computational representation, 
one is unable to reach the goal from 
Levels 2 and 3. Another example in Fig. 4 shows why we need two sub-levels in learning target selection. 
For the given character (here is {\it Albert Einstein}), one does need a linguistic representation 
to describe ``({\it un}){\it likeness''} \cite{Brennan} between the original image and caricature image. Only when 
a linguistic representation
is well defined, is a computational measure of similarity {\it possibly proper}
in caricature learning. The meaning of {\it possibly proper} is due to the difficulty in the following definition. 

\begin{figure}[htb]
    \centering{
    \hspace*{2cm} \includegraphics[width=50mm]{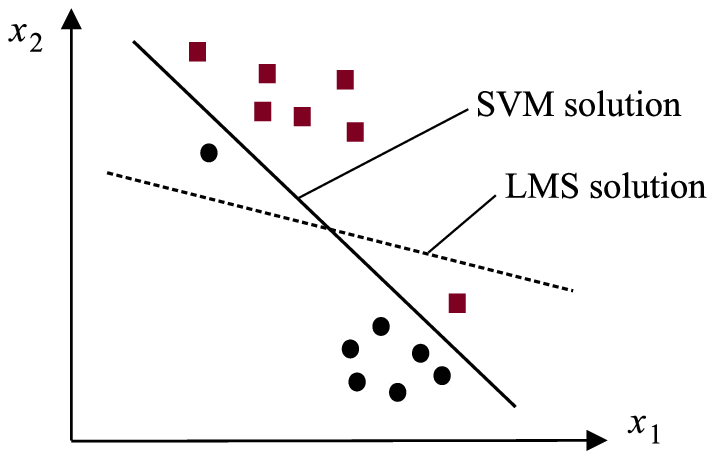}  \newline
    \small Fig. 3. Learning target selection within linearly separated dataset. 
    \newline (after \cite{Duda} on Fig. 5-17). Black Circle = Class 1, Ruby Square = Class 2.
}
\end{figure}

\begin{figure}[htb]
    \centering{
    \hspace*{2cm} \includegraphics[width=60mm]{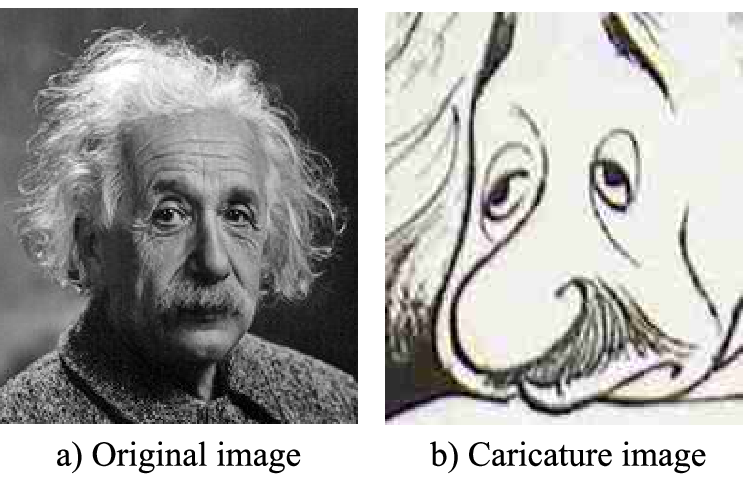}  \newline
    \small Fig. 4. Example of {\it ``What to learn?'' } and a need of defining a linguistic representation of 
    similarity for the given character. a) Original image ({\url{http://en.wikipedia.org/wiki/Albert_Einstein}}).
    b) Caricature image drawn 
    \newline by A. Hirschfeld ({\url{http://www.georgejgoodstadt.com/goodstadt/hirschfeld.dca}}).
}
\end{figure}

\textbf{Definition 5:} 
{\it ``Semantic gap'' }
is a difference between the two sets of representations. The gap can be linked by two
ways, namely, a {\it direct way} for describing
a connection from linguistic representation to computational representation, and 
an {\it inverse way} for a connection opposite to the direct one.

In this paper, I extend the definition of the gap in \cite{Smeulders} by distinguishing two ways.
The gap reflects one of the critical difficulties in machine learning.
For the direct-way study, the difficulty source mostly comes from {\it ambiguity} 
and {\it subjectivity} of the linguistic representation (say, on {\it mental} entity), which will lead to an {\it ill-defined}
problem. While sharing the same problem, an inverse-way study will introduce an extra challenge called 
{\it ill-posed}  problem, in which there is no {\it unique} solution (say, from a 2D image to 3D objects).    

Up to now, we have missed much studies on {\it learning target selection} 
if comparing with a study of {\it feature selection}. When {\it ``What to learn?''} is the most 
primary problem in machine learning, we do need a systematic, or comparative, 
study on this subject. The investigations from \cite{Rubinstein,Ng}
into {\it discriminative} and {\it generative} models confirm the importance of learning target selection 
in the vein of computational representation. From the investigations, one can identify 
the advantages and disadvantages of each model for applications. 
A better machine gaining the benefits from both models is developed \cite{Bishop}.
Furthermore, the subject of {\it ``What to learn?''} will provide a strong driving force to machine 
learning study in seeking {\it ``the fundamental laws that govern all learning processes''} \cite{Mitchell}. 

Take a decision rule about {\it ``Less costs more''} 
\footnote{
This rule is translated from Chinese saying, 
\begin{CJK}{UTF8}{bsmi}
{\it `` 物以稀為貴''} in Pinyin {\it ``Wu Yi Xi Wei Gui''}.
\end{CJK}
The translation is modified from the English phase {\it ``Less is more''} which usually describes 
{\it simplicity} in design. 
} 
for example. Generally, Chinese people classify object's values according to this rule. 
In {\it Big Data} processing, the useful information, 
that often belongs to a {\it minority class}, is extracted from massive datasets. 
While an English idiom describes it as {\it ``Finding a needle in a haystack''}, 
the Chinese saying refers to {\it ``Searching a needle in a sea} (\begin{CJK}{UTF8}{bsmi}\small{大海撈針})\end{CJK}''.
Users may consider that an error from a {\it minority class} will cost heavier than that from 
a {\it majority class} in their searching practices. This consideration will derive a decision rule 
like {\it ``Less costs more'' }. 
The rule will be one of the important strategies in Big Data processing.
Two questions can be given to the example. 
What is the {\it mathematical principle} (or {\it fundamental law}) 
for supporting the decision rule of {\it ``Less costs more''}? 
Is it a Bayesian rule? Machine learning study does need to answer the questions. 

\section{Information Theoretical Learning}\label{sec:Third}

\noindent Shannon introduced  {\it ``entropy''} concept as the basis of information theory \cite{Shannon}:
\begin{equation}
 H(Y) =  - \sum\limits_y {p(y)\log _2 p(y)},
\end{equation}
where $Y$ is a discrete random variable with {\it probability mass
function} $p(y)$. Entropy is an expression of {\it disorder} to the information. 
From this basic concept, the other {\it information measures} (or {\it entropy functions}) can be
formed (Table 1), where $p(t, y)$ is the {\it joint distribution} for the target
random variable $T$ and prediction random variable $Y$, and $p(t)$ and $p(y)$ are called
{\it marginal distributions}. 
We call them {\it measures} because some of them do not satisfy the {\it metric} properties fully, like
$KL$ {\it divergence} (asymmetric). 
More other measures from information theory can be listed as learning criteria, but
the measures in Table 1 are more common and sufficiently meaningful for the present discussion. 

\begin{table}[h]\footnotesize
  \caption{Some information formulas and their properties as learning measures.}
\begin{tabular}{ |l|l|l|l| }
  \hline
  Name & Formula & (Dis)similarity & (A)symmetry \\ \hline
  Joint Information & $ H(T,Y) = -\sum\limits_t \sum\limits_y {p(t,y)\log _2 {p
(t,y)}} $ & Inapplicable & Symmetry \\ \hline
  Mutual Information & $ I(T,Y) = \sum\limits_t \sum\limits_y {p(t,y)\log _2 \frac{{p
(t,y)}}{{p(t)p(y)}}} $ & Similarity & Symmetry \\ \hline
  Conditional Entropy & $ H(Y|T) = -\sum\limits_t \sum\limits_y {p(t,y)\log _2 {p
(y|t)}} $ & Dissimilarity & Asymmetry \\ \hline
  Cross Entropy & $ H(T;Y) = - \sum\limits_z {p_t (z)\log _2 {p_y
(z)}}  $ & Dissimilarity & Asymmetry \\ \hline
  KL Divergence & $ KL(T,Y) = \sum\limits_z {p_t (z)\log _2 \frac{{p_t
(z)}}{{p_y (z)}}} $& Dissimilarity & Asymmetry \\ \hline
\end{tabular}    
\end{table} 

We can divid the learning machines, in view of {\it ``mathematical principles''}, within two groups. 
One group is designed based on the empirical formulas, like {\it error rate} 
or {\it bound}, {\it cost} (or {\it risk}), {\it utility},  
or {\it classification margins}. The other is on information theory \cite{Principe,Yao}. 
Therefore, a systematic study seems 
necessary to answer the two basic questions below \cite{Hu2008}:

\begin{enumerate}
\item [Q1:] When one of the principal tasks in machine learning is to process data, can we apply entropy
or information measures as a generic learning target for dealing with uncertainty of data in machine learning?
\item [Q2:] What are the relations between information learning criteria and empirical learning criteria, 
and the advantages and limitations in using information learning criteria?
\end{enumerate}

Regarding to the first question, Watanabe \cite{Watanabe,Watanabe1981} proposed that
{\it ``learning is an entropy-decreasing process''} and 
pattern recognition is {\it ``a quest for minimum entropy''}.
The principle behind entropy criteria is to transform disordered data into ordered one (or {\it pattern}).
Watanabe seems to be the first 
{\it ``to cast the problems of learning in terms of minimizing properly defined entropy functions''} \cite{Safavian},
and throws a brilliant light on the learning target selection in machine learning. 

In 1988, Zellner theoretically proved that Bayesian theorem can be derived from the optimal 
information processing rule \cite{Zellner}. This study presents a novel, yet important, finding
that Bayesian theory is rooted on information and optimization concepts. 
Another significant contribution is given by Principe and his collaborators
\cite{Principe2000,Principe} for the proposal of Information 
Theoretical Learning (\textbf{ITL}) as a generic learning target in machine learning. We consider 
ITL will stimulate us to develop new learning machines as well as {\it ``theoretical interpretations''} of
learning mechanisms. 
Take again the example of the decision rule about {\it ``Less costs more''}.
Hu \cite{Hu2014} demonstrates theoretically that Bayesian principle is unable to support the rule. 
When a minority class approximates to a zero population, Bayesian classifiers will tend to
misclassify the minority class completely.  
The numerical studies \cite{Hu2014,Zhang} show that {\it mutual information} provides 
positive examples to the rule. The classifiers based on  
{\it mutual information} are able to protect a minority class and automatically balance 
the error types and reject types in terms of population ratios of classes. Theses studies 
reveal a possible mathematical interpretation of learning mechanism behind the rule. 

\section{(Dis)similarity Measures in Machine Learning}\label{sec:Fourth}

\noindent When mutual information describes similarity between two variables, the other information
measures in Table 1 are applied in a sense of dissimilarity. For a better understanding of them,
their graphic relations are shown in Fig. 5. If we consider the variable $T$ provides a ground truth
statistically 
(that is, $p(t)=(p_1, ..., p_m)$ with the {\it population rate} $p_i (i=1,...,m)$ is known and fixed), 
its entropy $H(T)$ will be the baseline in learning. 
In other words, when the following relations hold,

\begin{eqnarray}
    I(T,Y)= H(T;Y)=H(Y;T)=H(Y)=H(T), ~ o r  \nonumber  \\ 
     KL(T,Y)=KL(Y,T)=H(T|Y)=H(Y|T)=0,
\end{eqnarray}
we call the measures {\it reach} the baseline of $H(T)$.

\begin{figure}[htb]
\begin{center}
    \centering{
    \hspace*{1.3cm} \includegraphics[width=80mm]{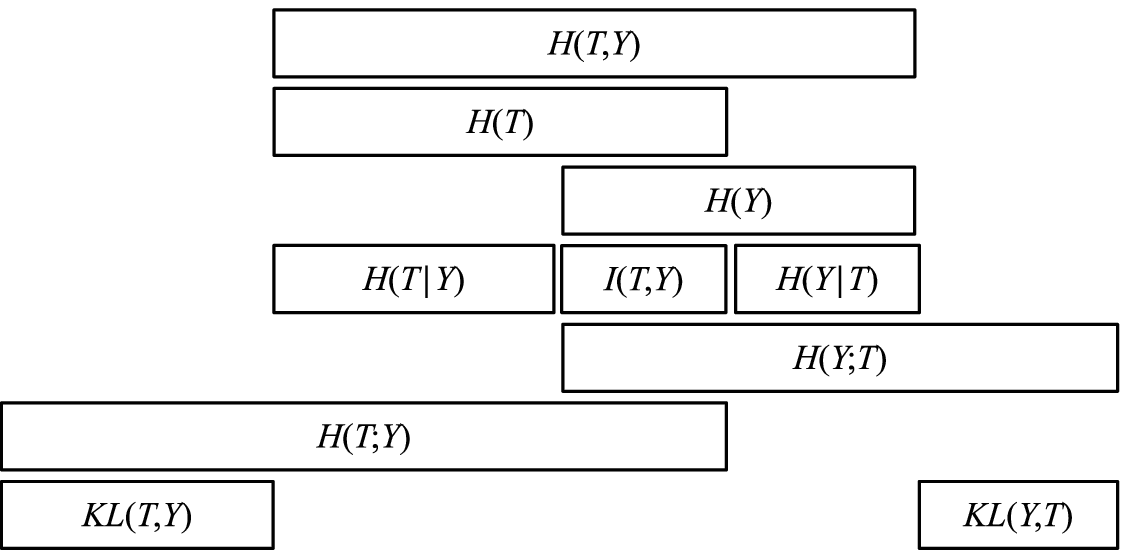}  \newline \newline
    \small Fig. 5. Graphic relations among joint information, mutual information, 
    \newline marginal information, conditional entropy, cross entropy and $KL$ divergences 
    \newline (modified based on \cite{Mackay} by including cross entropy 
and $KL$ divergences).
}
\end{center}
\end{figure}

Based on the study in \cite{Hu2012}, further relations are illustrated
in Fig. 6 between exact classifications and the information measures. 
We apply the notations of $E, Rej, A, CR$ for the {\it error, reject, accuracy, and correct recognition rates}, respectively.
Their relations are given by:
\begin{eqnarray}
    CR+E+Rej=1,   \nonumber  \\ 
    A= \frac{CR} {CR+E}.
\end{eqnarray}
The form of $\{ y_k \} =\{ t_k \} $ in Fig. 6 describes an equality between the label variables in every samples.  
For a finite dataset, the empirical forms
should be used for representing the distributions and measures \cite{Hu2012}.  
Note that the link using  ``$\leftrightarrow$'' indicates a {\it two-way} connection for equivalent relations,
and  ``$\rightarrow$'' for a {\it one-way} connection. Three important aspects can be observed from 
Fig. 6:

\begin{itemize}
\item[I.] The {\it necessary} condition of exact classifications is that all the information measures reach the
baseline of $H(T)$.
\item[II.] When an information measure reaches the baseline of $H(T)$, it does not {\it sufficiently} indicate an
exact classification. 
\item[III.] The different locations of one-way connections result in the interpretations {\it why} and {\it where} 
the sufficient condition exists.  
\end{itemize}

\begin{figure}[htb]
    \centering{
    \hspace*{0.4cm} \includegraphics[trim = 0mm 0mm 26mm 0mm, clip, width=95mm]{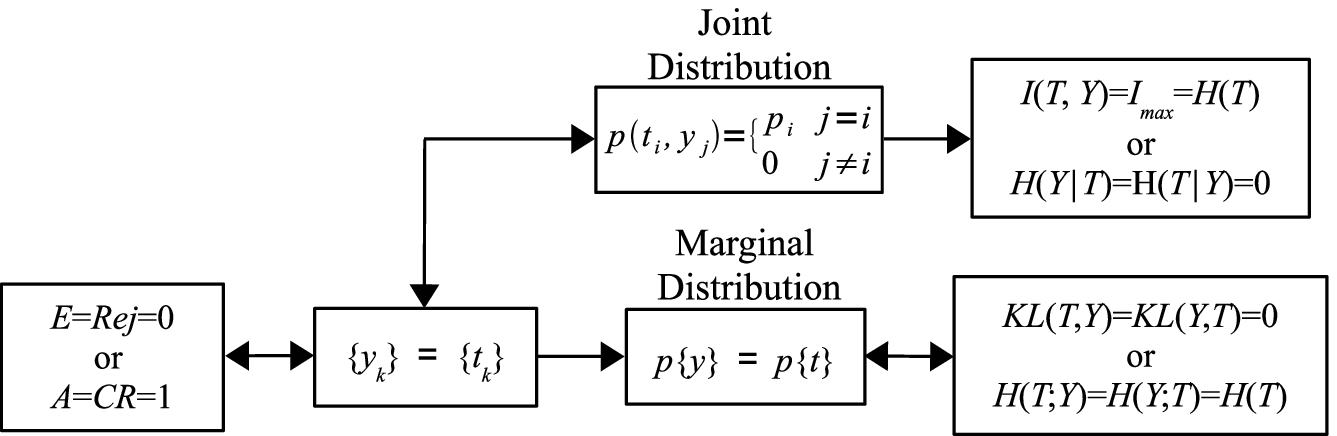}  \newline \newline
    \small Fig. 6. Relations between exact classifications and mutual information, 
    \newline conditional entropy, cross entropy and $KL$ divergences. 
}
\end{figure}

Although Fig. 6 only shows the relations to the information measures listed in Table 1 for the classification problems, 
its observations may extend to other information measures as well as to the other problems, like clustering,
feature selection/extraction, image registrations, etc. When we consider machine learning or pattern recognition to be a 
process of data in a {\it similarity} sense (any dissimilarity measure can be
transformed into similarity one \cite{Hu2012}), one important theorem exists to describe their relations.

\textbf{Theorem 1}. Generally, there is no {\it one-to-one} correspondence between the
empirically-defined similarity measures and information measures. 

The proof is neglected in this paper, but it can be given based on the study 
of bounds between entropy and error 
(cf. \cite{Hu2013} and references therein). 
The significance of Theorem 1 implies that an optimization of 
information measure may not guarantee to achieve 
an optimization of the empirically-defined similarity measure.

\section{Final remarks}\label{sec:Final}

\noindent Machine learning can be exploited with different perspectives depending on study goals of researchers. 
For a deep understanding of ourself on the learning mechanisms mathematically, 
we can take learning machines as {\it human's extended sensory perception}. 
This paper stresses on identifying the primary problem in machine learning from a novel perspective.
I define it as {\it ``What to learn?''} or {\it ``learning target selection''}. Furthermore,
two sets of representations are specified, namely, {\it ``linguistic representation''} and 
{\it ``computational representation''}. While a wide variety of computational representations  
have been reported in learning targets, we can argue that if there exists a unified, yet fundamental, principle
behind them. Towards this purpose, this paper extends the Watanabe's proposal \cite{Watanabe,Watanabe1981} and the
studies from Zellner \cite{Zellner} and Principe \cite{Principe}
to a {\it ``conjecture of learning target selection''} in the following descriptions. 

\textbf{Conjecture 1}. In a machine learning study, all computational representations of learning target(s)
can be interpreted, or described, by optimization of entropy function(s).

I expect that the proposal of the conjecture above will 
provide a new driving force 
not only for seeking fundamental laws governing all learning processes \cite{Mitchell}
but also for developing improved learning machines \cite{He} in various applications. 

\section*{Acknowledgment}\label{sec:Acknowledgment}

This work is supported in part by NSFC(No. 61273196).

\end{document}